\documentclass[12pt,epsfig]{iopart}
\usepackage{epsfig}

\begin{document}

\title{On the quantum analogue of Galileo's leaning tower experiment}

\author{Md. Manirul Ali\dag\, A. S. Majumdar\dag \footnote[3]{archan@bose.res.in}\, Dipankar Home\ddag\ and Alok Kumar Pan\ddag}

\address{\dag\ S. N. Bose National Centre for Basic Sciences, Block JD,
Sector III, Salt Lake, Calcutta 700098, India}

\address{\ddag\ Department of Physics, Bose Institute, Calcutta
700009, India}

\begin{abstract}

The quantum analogue of Galileo's leaning tower experiment is revisited
using wave packets evolving under the gravitational potential. We first
calculate the position
detection probabilities for particles projected upwards against gravity
around the classical turning point and also around the point of initial
projection, which exhibit mass dependence at both these points. 
We then compute the mean arrival time of freely 
falling particles using the quantum probability current, which also 
turns out to be mass dependent. The mass dependence of 
both the position detection probabilities and the mean arrival time vanish 
in the limit of large mass. Thus, compatibility between the weak equivalence 
principle and quantum mechanics is recovered in the macroscopic limit of 
the latter.

PACS number(s): 03.65.Ta,04.20.Cv

\end{abstract}

\maketitle

\section{Introduction}

As a consequence of the equality of gravitational and inertial mass, all
classical test bodies fall with an equal acceleration independently
of their mass or constituent in a gravitational field. Historically, the 
first experimental study to probe this feature was conceived by Galileo with 
test bodies in free fall from the leaning tower of Pisa\cite{galileo}. In 
modern 
times several tests have been performed with pendula or torsion balances 
leading 
to extremely accurate confirmations of the equality of gravitational and 
inertial 
masses\cite{will}. Though most of these schemes consider only classical test 
bodies, 
there exist indications about the validity of the equality of 
gravitational and 
intertial masses even for quantum mechanical particles using the 
gravity-induced 
interference experiments\cite{colella,chu}. The universal character of the 
law of 
gravitation, however, has a much richer structure than the above equality, as 
embodied in the principle of equivalence in its various versions.

There are three statements of the equivalence principle which are
equivalent according to classical physics but are logically distinct. 
Holland\cite{peter} emphasized the importance of separating them clearly in 
order to discuss their quantum analogues: (i) {\it Inertial mass is equal to
Gravitational mass; $m_i=m_g=m$}. As mentioned earlier, the compatibility 
of this
equality with quantum mechanics has been verified in several 
experiments\cite{colella,chu}. (ii) {\it With respect to the mechanical 
motion 
of particles, a state of rest in a sufficiently weak, homogeneous gravitational
field is physically indistinguishable from a state of uniform acceleration in a
gravity-free space}. A natural quantum analogue of this statement 
is\cite{bonse}: 
``The laws of physics are the same in a frame with gravitational potential
$V=-mgz$ as in a corresponding frame lacking this potential but having
a uniform acceleration $g$ instead''. This can verified in quantum mechanics 
by 
transforming the Schr${\ddot o}$dinger wave function for a quantum particle in
a gravitational 
potential to that in an accelerated frame lacking this potential\cite{green3}. 
Predictions of the Schr${\ddot o}$dinger equation in a noninertial frame 
have been 
shown to be experimentally observed\cite{bonse}. (iii) {\it All sufficiently 
small 
test bodies fall freely with an equal acceleration independently of their 
mass or 
constituent in a gravitational field}. To obtain its quantum analogue this 
statement 
might be replaced by some principle such as the following\cite{peter}: ``The 
results of 
experiments in an external potential comprising just a (sufficiently weak, 
homogeneous) 
gravitational field, as determined by the wavefunction, are independent 
of the mass of 
the system''. The status of this last version of the 
equivalence 
principle for quantum mechanical entities is the subject of investigation of 
the present 
paper. We shall henceforth call the quantum analogue of version (iii) as the 
weak equivalence 
principle of 
quantum mechanics ({\it WEQ}). 

The compatibility between {\it WEQ} and quantum 
mechanics 
is an interesting issue which is yet to be completely settled.
This issue was  
elaborately discussed by Greenberger\cite{green1}.
Evidence supporting the violation of 
{\it WEQ} already 
exists in interference phenomena associated with the gravitational potential 
in neutron 
and atomic interferometry experiments\cite{colella,chu} where the observable 
interference patterns are mass dependent. Further, at the theoretical level, 
on applying 
quantum mechanics to the problem of a particle bound in an external 
gravitational
potential it is seen that the radii, frequencies and  binding energy depend 
on the mass 
of the bound particle\cite{green3,green1,green2}. The possibility of quantum 
violation of {\it WEQ}
 is also discussed in a number of other papers, for 
instance 
using neutrino mass oscillations in a gravitational potential\cite{vep}.

Recently, Davies\cite{paul} has 
provided a particular quantum mechanical treatment of the violation of 
{\it WEQ} for a quantum particle whose time of flight is proposed to be 
measured by a 
model quantum clock\cite{peres}. This model of quantum clock actually measures 
the phase change of 
the wave 
function during the particle's passage through 
a specified spatial region. In this 
treatment, 
Davies considered a variant of the simple Galileo experiment where particles 
of different 
mass are projected vertically in a uniform gravitational field. 
Quantum 
particles are able to tunnel into the classically forbidden region beyond
the classical turning point and the tunneling depth 
depends on the mass. One might therefore expect a small but significant 
mass-dependent 
``quantum delay'' in the return time. Such a delay would represent a violation
 of {\it WEQ}.
Using the concept of the Peres clock\cite{peres} the time of flight is 
calculated 
from the {\it stationary state} wave function for the quantum particle moving 
in a gravitational 
potential. However, this violation is {\it not} found far away from the 
classical turning point 
of the particle trajectory. Within a distance of roughly one de Broglie wave 
length from the
classical turning point there are significant quantum corrections to the 
turn-around time (i.e., the time taken by the particle to reach its
maximum height),
including the possibility of a mass-dependent delay due to the penetration of 
the classically
forbidden region by the evanescent part of the wavefunction. Thus, this 
quantum ``smearing'' 
of the {\it WEQ} is restricted to distances within the usual 
position uncertainty
of a quantum particle. 

In another relevant gedanken experimental scheme Viola and Onofrio\cite{viola} 
have studied the free fall of a quantum test particle in a uniform 
gravitational field. 
Using Ehrenfest's theorem for obtaining the average time of flight for a test 
mass,  if one takes gravitational mass to be equal to the inertial 
mass then the 
mean time taken by the particle to traverse a distance $H$ under free fall 
is $\langle t \rangle=\sqrt{{2 H}/g}$ which is exactly equal to the classical 
result. Viola and Onofrio made a rough 
estimate of the fluctuations around this mean value using a 
semiclassical 
approach with the initial wave function taken as a Schr${\ddot o}$dinger cat 
state. This 
fluctuation around the mean time of flight was shown to be dependent on the 
mass of the 
particle. However, one may note that the very definition of the time of 
flight or arrival time 
of a quantum 
particle is the subject of much debate, and there exists no unique
or unambiguous definition that is universally applicable
and also empirically well-tested\cite{reviews}.

As a sequel to these works by Viola and Onofrio\cite{viola} and 
Davies\cite{paul}, we study the issue of violation of {\it WEQ} in the present 
paper from a different perspective. Note that the 
gravitational equivalence principle has been historically formulated
at the level of single particles, which is quite appropriate within the
domain of classical mechanics. However, the formulation of the
quantum counterpart is experimentally verifiable only at the 
level of an ensemble evolving
through Schrodinger dynamics. Following this line of argument, it
seems that for a quantum-classical comparison to be meaningful, even the 
classical results have to be stated in the framework of a distribution
of particles undergoing a classical dynamical evolution\cite{ali2}.
To this end
we consider an ensemble of identical 
quantum particles 
represented by a Gaussian wave packet which evolves under the gravitational 
potential. We make use of the quantum 
probability current 
in computing 
the mean arrival time for a wave packet under free fall. The probablility
current approach\cite{current} towards calculating the mean arrival time 
for an ensemble
of quantum particles is conceptually sound and also well suited for 
our present investigation of the violation of {\it WEQ}.

The plan of the paper is as follows.
In the next section we compute the {\it position detection probability} for 
atomic and 
molecular mass particles represented by a Gaussian wave packet that is
projected upwards 
against gravity around two different points; one around the classical turning 
point and another 
around a region of the initial projection point after it returns 
back. We show an explicit {\it mass dependence} of the position probability 
computed around both 
these points, thus indicating violation of  
{\it WEQ} not only 
at the {\it turning point} of the classical trajectory, but also 
{\it far away} from it around the 
{\it initial projection point}. We then compute 
the mean arrival time for a wave packet under free fall in Section III. 
Here we 
consider the case 
when the particles are dropped from a height with zero initial velocity.
We observe an explicit 
{\it mass dependence} of the {\it mean arrival time} at an arbitrary 
detector location indicating once
again the manifest violation of {\it WEQ}. Another issue of 
interest as discussed by Greenberger\cite{green2} is to understand whether 
compatibility 
with {\it WEQ} is recovered in the 
macroscopic limit of 
quantum mechanics. We show that using the quantum probability current 
approach of 
obtaining the mean arrival time\cite{current} of an ensemble of particles,
the validity of {\it WEQ}
emerges smoothly in the limit of large mass.
We conclude with a brief summary of our results in Section IV highlighting the 
key differences of
our approach with the earlier works.

\section{Mass dependence of position detection probabilities}

A beam of quantum particles with an initial Gaussian distribution is considered
to be projected upwards against gravity. Subsequently, the position probability
distribution is calculated within an arbitrary region either around the 
classical 
turning point of the potential $V=m_g g z$ or away from the turning point 
around 
the region from where the particles were projected. Such an observable quantity
turns out to be mass dependent, as seen below.

Let us consider particles of different inertial masses that are thrown upward
against gravity with the same initial mean position and mean velocity. The
initial states of the quantum particles can be represented by a one 
dimensional Gaussian wave
function given by                                                        
\begin{eqnarray}
\psi^j (z, t=0)= \left(2\pi \sigma^{2}_{0}\right)^{-1/4} \exp \left(i k^j z 
\right)
\exp \left( -\frac{z^{2}}{4\sigma^{2}_{0}}\right)
\end{eqnarray}                 
peaked at $z=0$ with the initial group velocity (defined for the above wave 
function
evolving through the Schrodinger equation as 
$u=(d\omega_j)/(dk_j)$ with $\omega_j$ and $k_j$ being the angular frequency 
and wave number, respectively, for the $j^{th}$ particle)  given by  
$u={\hbar k^j}/{m_i^j}$, 
where $m_i^j$ is the inertial mass of the $j^{th}$ particle.

In order to perform an ideal free fall experiment for quantum particles having
different inertial masses $m_i^{1}$, $m_i^{2}$,.. etc. (with suffix {\it i}
representing the inertial mass, and with $m_i^{1} \ne m_i^{2}$ etc.), one has
to specify an initial preparation in such a way that any difference
in the motion during the free fall must be ascribed to the effect of gravity.
Now, within the classical Hamilton picture the Galileian prescription for 
initial
positions and velocities fixes the ratio between the initial momenta in a 
well-defined
way, i.e., $p_0^{1}/p_0^{2}=m_i^{1}/m_i^{2}$, etc. Following Ref.\cite{viola},
 we
extend such a prescription to the quantum case, of course keeping in mind
that the Heisenberg uncertainty principle forbids the simultaneous definition
of the initial position and momentum for each particle. 
If $\psi_1$ and $\psi_2$
denote the initial wave functions for particles 1 and 2 in the 
Schr${\ddot o}$dinger
picture, the quantum analogue of the situation can be achieved by 
stipulating the
conditions
\begin{eqnarray}
{\langle {\widehat z}\rangle}_{\psi^1}={\langle {\widehat z}\rangle}_{\psi^2}=0,
~~~~ \frac{{\langle {\widehat p_z}\rangle}_{\psi^1}}{m_i^{1}}=
\frac{{\langle {\widehat p_z}\rangle}_{\psi^2}}{m_i^{2}} \equiv u
\end{eqnarray}
where ${\langle {\widehat z}\rangle}_{\psi}$ and
${\langle {\widehat p_z}\rangle}_{\psi}$ denote the expectation values
for position and momentum operators, respectively (confining to a one 
dimensional representation
along the vertical $z$ direction). The probabilistic interpretation
underlying quantum mechanics allows us only to speak of probability 
distributions,
for instance, characterized by {\it mean} initial conditions such as Eq.(2), as
opposed to the sharply-defined values for the relevant classical observables.
                                                                    
With the above prescription one can consider the time evolution of the
initial state under the potential $V=m_g^j g z$, where $m_g^j$ is the 
gravitational mass of the $j^{th}$ particle. At any subsequent
time $t$ the Schr${\ddot o}$dinger time evolved 
wave function $ \psi^j \left( z,t\right)$ is given by
\begin{eqnarray}
\nonumber
&&\psi^j \left( z,t\right)=\left(2\pi s^{2}_{t}\right)^{-1/4}
\exp \left[ \frac{ \left( z- ut + ({m_g^j}/{m_i^j}) \frac{1}{2}g
t^2\right)^2}{4s_{t}\sigma_{0}} \right]\\
\nonumber
&&\times~ \exp \left[ i({m_i^j}/{\hbar}) \left\{\left(u- ({m_g^j}/{m_i^j})gt\right)
\left(z-ut/2\right) \right\} \right]\\
&&\times~ \exp\left[ i({m_i^j}/{\hbar}) \left\{-({m_g^j}/{m_i^j})^2 \frac{1}{6}g^2 t^3 \right\} \right]
\end{eqnarray}
where $s_{t}=\sigma_{0}\left(1+i\hbar t/2{m_i^j}\sigma_{0}^{2}\right)$. We
see even if one takes $m_i^j=m_g^j$, i.e., equates the inertial mass with the
gravitational mass, the observable position probability density $\left|\psi^j
\left( z,t\right) \right|^{2}$ will have an explicit mass dependence
\begin{eqnarray}
|\psi^j \left( z,t\right)|^2=\left(2\pi \sigma^2 \right)^{-1/2}
\exp \left[-\frac{\left(z-ut+ \frac{1}{2} g t^2\right)^2}{2 \sigma^2} \right]
\end{eqnarray}
coming from the spreading of the wave packet given by  
$\sigma=\sigma_{0}\left(1+\hbar^{2}t^{2}/4{m_i^j}^{2}\sigma_{0}^{4}
\right)^{1/2}$ which is mass dependent.

\begin{table}
\centering
\caption{\label{tab:table1} 
Mass dependence of the probability at the initial projection point.
We take $u=10^3$ cm/sec, $\sigma_0=10^{-3}$ cm, $\epsilon=\sigma_0$, $t=t_2 = 
2u/g$ sec.}
\vskip 0.3cm
\begin{tabular}{|c|c|c|}
\hline
$System$&$Mass (m_i^j)$&$Probability$\\
$~$&$in (a.m.u)$&$P_1(m_i^j)$\\
\hline
$H$&1.00&0.0012\\
$H_2$&2.00&0.0024\\
$Li$&6.94&0.0085\\
$Be$&9.01&0.0111\\
$C$&12.01&0.0148\\
$Ag$&107.87&0.1305\\
$C_{60}$&720.00&0.5428\\
$\mathrm{protein~~molecule}$&$7.2 \times 10^4$&0.6826\\
$\mathrm{heavier~~molecule}$&$7.2 \times 10^7$&0.6826\\
\hline
\end{tabular}
\end{table}

The peak of the wave packet follows the classical trajectory and it has a
turning point at the time $t=t_1=u/g$ at $z=z_c=u t_1$. At a later time
$t=t_2=2u/g$, when the peak of the wave packet comes back to its initial
position $z=0$, if we compute the probability of finding particles 
$P_1({m_i^j})$
within a very narrow region $(-\epsilon~~ to~~ +\epsilon)$ around this point 
$z=0$ 
then that probability is found to be a {\it function of mass} and is given by
\begin{eqnarray}
P_1\left({m_i^j}\right)=\int _{-\epsilon }^{+\epsilon }|\psi^j (z,t_{2})|^{2}dz
\end{eqnarray}
This effect of the {\it mass dependence} of the probability occurs essentially
because
the spreading of the wave packet under gravitational potential is different for
particles of different masses. We explicitly estimate  this effect for 
different molecular mass particles. A different set of mass dependent 
probabilities
$P_1\left({m_i^j}\right)$ may be 
obtained by taking a different value of the width $\sigma_0$ of the 
initial wave packet.
In the Table-\ref{tab:table1} it is shown numerically how the
probability of finding the particles $P_1\left({m_i^j}\right)$
around the mean initial projection point ($z=0$) changes with the
variation of mass for an initial Gaussian position distribution.
We note that for further increase in mass of the particle beyond that of a 
protein molecule, the change in the probablity
$P_{1}(m_{i}^{j})$ gets negligbly small, or in other words the mass 
dependence of the probability gets saturated.

\begin{table}
\centering
\caption{\label{tab:table2}   
Mass dependence of the probability at the turning point.
We take $u=10^{3}$ cm/sec, $\sigma_0=10^{-3}$ cm, $\epsilon$=$\sigma_{0}$, 
$t=t_{1}=u/g$ sec.}
\vskip 0.3cm
\begin{tabular}{|c|c|c|}
\hline
$System$&$Mass (m_i^j)$&$Probability$\\
$~$&$in (a.m.u)$&$P_2(m_i^j)$\\
\hline
$H$&1.00&0.0024\\
$H_2$&2.00&0.0049\\
$Li$&6.94&0.0171\\
$Be$&9.01&0.0222\\
$C$&12.01&0.0296\\
$Ag$&107.87&0.2522\\
$C_{60}$&720.00&0.7277\\
$\mathrm{protein~~molecule}$&$7.2 \times 10^4$&0.7978\\
$\mathrm{heavier~~molecule}$&$7.2 \times 10^7$&0.7978\\
\hline
\end{tabular}
\end{table}

We then compute the probability of finding particles 
$P_{2}\left( m_{i}^{j}\right)$ 
at $t=t_{1}=u/g$ within a very narrow detector region 
$\left(-\epsilon~~to~~+\epsilon \right)$ around a point which is the 
classical turning point $ z=z_{c}=ut_{1}$ for the
particle. $P_{2}\left( m_{i}^{j}\right)$ is also a function of mass 
and is given by
\begin{eqnarray}
P_{2}(m_{i}^{j})=\int_{-\epsilon }^{+\epsilon }|\psi^j (z,t_{1})|^{2}dx
\end{eqnarray}
In the Table-\ref{tab:table2} it is shown numerically how the 
probability of finding the particles $P_{2}(m_{i}^{j})$ around the 
classical 
turning point changes with the variation of mass for a intial Gaussian 
position distribution.
As in the previous case, we again find  that the mass-dependence of the 
probablity $P_{2}(m_{i}^{j})$ for finding
the particle gets saturated in the limit of large
mass. 

The question of the quantum-classical correspondence\cite{home} 
could be elaborated
further within the present context by constructing a suitable classical phase
space distribution matching with the initial quantum distribution. 
It may be interesting to note that
if one were to work with a classical ensemble of particles with an
initial phase space distribution taken as the product of two Gaussian
functions matching the initial position distribution $|\psi(z,0)|^2$
and its fourier transform (say, $|\phi(p,0)|^2$ representing the initial
momentum distribution), essentially the same results attributed to
ensemble spread are obtained through the classical Liouville evolution
for Gaussian distributions\cite{ali2}.  Note also that within the
present context the use of the Wigner function does not lead to any
new insights since for the linear gravitational potential the Wigner 
function reproduces classical results.

\section{Mass dependence of mean arrival time and the classical limit}

Now let us pose the problem in a different way. We consider the quantum 
particle
prepared in the initial state given by Eq.(1) satisfying Eq.(2) and with $u=0$.
The particle is subjected to free fall under gravity. We then ask the question
as to when does the
quantum particle reach a detector located at $z=Z$. In classical mechanics, 
a particle follows a definite trajectory; hence the time at which a particle 
reaches a given location is a well defined concept. On the other hand, in 
standard
quantum mechanics, the meaning of arrival time has remained rather obscure. 
There exists an extensive literature on the treatment of arrival time 
distribution 
in quantum mechanics\cite{reviews}. One possible internally consistent  
approach of  
defining the arrival time probability distribution is through the quantum 
probability 
current\cite{current} which we employ in the present investigation. 
The probability current approach for computation of the mean arrival time of
 a quantum 
ensemble not only provides an unambiguous definition of arrival time at the 
quantum mechanical 
level\cite{current,holland,ali}, but also addresses the issue of obtaining 
the proper 
classical limit of the time of flight of massive quantum particles\cite{ali2}.

It is relevant to observe here that though the Schr${\ddot o}$dinger 
probability 
current is \emph{not}
uniquely defined within nonrelativistic quantum mechanics, but for, say,
particles with spin-$1/2$, it has been shown by Holland\cite{holland} by 
taking the nonrelativistic limit of the Dirac probability current,
that the quantum probability current
contains a term that is spin dependent.
The arrival time distribution is then uniquely formulated using
the probability current obtained by taking the nonrelativistic limit
of the corresponding relativistic current.
It was shown using the explicit example of a Gaussian wave packet
that the spin-dependence of the probability current leads  to the 
spin-dependence
of the mean arrival time for free particles\cite{ali}. However, for the case of
massive spin-0 particles it has been shown recently by taking the 
non-relativistic
limit of  Kemmer equation\cite{kemmer} that the unique probability current is 
given by the Schr${\ddot o}$dinger current\cite{baere}. Hence, the 
Schr${\ddot o}$dinger probability current density can be used to define 
a precise and logically consistent arrival time distribution 
for spin-0 quantum particles, that is relevant for the 
present analysis.  

The expression for the Schr${\ddot o}$dinger  probability current density 
$J(Z,t)$
at the detector location $z=Z$ for the time evolved state is calculated using 
the 
initial state prepared in the Gaussian form given by Eq.(1) and 
satisfying Eq.(2). The particle falls 
freely under gravity along $-{\widehat z}$ direction from 
the initial 
peak position at $z=0$ with 
$u=0$ and $J(Z,t)$ is given by
\begin{eqnarray}
J(Z,t)=\rho (Z,t)\hskip 0.2cm v(Z,t)
\end{eqnarray}
where
\begin{eqnarray}
\rho (Z,t)=(2\pi \sigma^{2})^{-1/2}\hskip 0.1cm exp\left[-\frac{(Z-\frac{1}
{2}gt^{2})^{2}}{2\sigma^{2}}\right]
\end{eqnarray}
and
\begin{eqnarray}
v(Z,t)=\left[ gt+\frac{\hbar^{2} t}{4{m_i^{\tiny j}}^{2}\sigma_{0}^{2}
\sigma^{2}}(Z-gt^{2}/2)\right]
\end{eqnarray}

Taking the modulus of the probability current density as determining the
arrival time distribution\cite{current}, the mean arrival time 
$\overline{\tau}$ at a
particular detector location is computed for an ensemble of particles with
an initial Gaussian position distribution {\it falling freely} under
gravity. Then this observable quantity $\overline{\tau}$ is given by
\begin{eqnarray}
\overline{\tau}\left(m_i^j\right)=\frac{\int_{0}^{\infty}
\left|J\left(Z,t\right) \right|t\, dt}{\int _{0}^{\infty}
\left| J\left(Z,t\right)\right|dt}
\end{eqnarray}
which is actually the first temporal moment of the modulus of the probability
current density.
Since $\sigma =|s_{t}|=\sigma_{0}\left(1+\hbar ^{2}t^{2}/4{m_i^{\tiny j}}^{2}
\sigma_{0}^{4}\right)^{1/2}$ is mass dependent, it is seen from Eqs.(7--9) 
that 
$ J\left( Z,t\right)$ is mass-dependent too. Hence the mean arrival time 
$\overline{\tau }$ calculated by using Eq.(10) for the Gaussian wave packets 
corresponding to different atomic mass particles falling freely under gravity 
is also mass dependent. 

In FIG.1, we depict the variation with mass of the mean arrival
time at a particular detector location for an ensemble of particles under free
fall. The initial conditions are taken as $\langle z \rangle_0 =0 $ 
and $\langle p \rangle_0 =0 $, where $\langle z \rangle_0$ and 
$\langle p \rangle_0$ 
are the position and momentum expectation values at $t=0$. One should note 
that though the integral in 
the numerator of Eq.(10) formally diverges, several techniques have 
been employed in the literature ensuring rapid fall off for the 
probability distributions asymptotically\cite{hahne}, so that convergent 
results are obtained
for the integrated arrival time. For our present purposes it is sufficient
to employ the  simple strategy of taking  
a cut-off ($t=T$) in the upper limit of the time integral with 
$T=\sqrt{2(Z+3 \sigma_T )/g}$
where $\sigma_T$ is the width of the wave packet at time $T$. Thus, 
our 
computations of the arrival time are valid up to the $3 \sigma$ level of 
spread in 
the wave function. 

\begin{figure}
\centering
\epsfxsize=5.0in\epsfysize=3.0in
\epsfbox{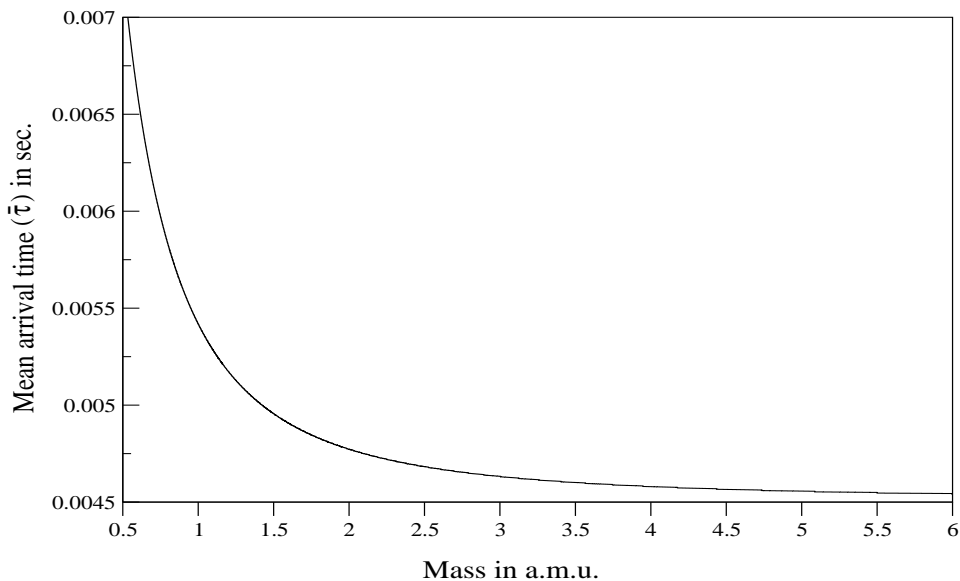}
\caption{\label{fig:epsart}
The variation of mean arrival time with
mass (in atomic mass unit) at a detector location $Z$ for an initial
Gausssian position
distribution. We take $\sigma_{0}=10^{-4}~~cm$,
$Z=10^{-2}~~cm$.}
\end{figure}

One can  see from FIG.1 that in the limit of large mass the mean arrival 
time 
$\overline{\tau}$ asymptotically approaches the classical result which is
mass independent. As was discussed by Greenberger\cite{green2},
the question as to whether
compatibility of the weak equivalence principle with quantum mechanics 
emerges in the 
classical limit is clouded by 
conceptual intricacies of obtaining the proper macroscopic limit of quantum 
mechanics. We see here again the probability current approach
offers an effective and consistent scheme for obtaining the macroscopic limit
of the arrival time distribution by continuously increasing the mass of the 
particle. We find that the classical value of mean arrival time is obtained
as the mass dependence vanishes in the limit of large mass. We are thus 
able to show that compatibility of the weak equivalence principle with
quantum mechanics emerges in a smooth manner in the macroscopic limit.

\section{Summary and conclusions}

To summarize, we have revisited a gedanken version of the quantum analogue of 
Galileo's 
leaning tower experiment with atomic and molecular mass wave packets falling 
freely under 
gravity. Our results of mass-dependence of the position detection 
probabilities and the 
arrival time distribution clearly indicate the manifest violation of the 
quantum analogue
\cite{peter} of the weak equivalence principle ({\it WEQ}) stated earler. 
Davies\cite{paul} 
provided a particular quantum mechanical treatment of the violation of 
{\it WEQ} using the concept of the Peres clock\cite{peres} where the time of 
flight is calculated from the 
stationary state wave function for the quantum particle moving in a 
gravitational potential. 
However, this violation was {\it not} found far away from the classical 
turning point of the 
particle trajectory and was restricted to distances within the usual
position uncertainty 
of the quantum particle. 
A semi-classical approach based on the Ehrenfest theorem yields the 
classical result for 
the average time of flight and mass dependence for fluctuations around the 
average\cite{viola}.
Our approach, on the other hand, is based on
the quantum probability current approach and leads to the mass 
dependence of the 
arrival time distribution computed around any position along the trajectory of 
the particles. The predicted violation of {\it WEQ} in this case is, 
in principle,
observable  for molecular mass particles.

We have further discussed the issue of compatibility of  {\it WEQ}
with the macroscopic limit of quantum mechanics\cite{green2}. For this purpose
it is essential to consider the evolution of an ensemble of particles
that we have done using a Gaussian wave packet. We see that the 
variation of the 
detection probability with mass disappears in the limit of large mass of the 
freely falling 
particles, as is expected for classical objects. This saturation of the 
detection probability 
is also reflected in the mean arrival time defined through the quantum 
probability current, 
which approaches the classical result in a continuous manner with the 
increase of mass. 
We have seen 
that the compatibility of {\it WEQ} with quantum 
mechanics can be
restored in the classical limit within this framework for particles falling 
freely under 
gravity. Our analysis has been carried out
using a minimum uncertainty Gaussian wave packet. Following our approach, 
it should be interesting
to investigate the issue of compatibility of the weak equivalence
principle with quantum mechanics in
the macroscopic limit for other types of Gaussian and 
non-Gaussian wave packets. 

Finally, we would like to re-emphasize that our approach of demonstrating the 
quantum violation of the weak equivalence principle is different from that 
of other examples in that using our 
scheme it should be possible to predict the specific 
mass range of
molecules where an explicit violation of {\it WEQ} may occur either
through the measurement of the position detection probabilities, or through
the mean arrival time. 
Our approach is capable of providing a precise prediction of the 
quantum violation of the weak equivalence principle in the relevant mass
ranges as one goes from the micro to macro limit, and is thus
amenable to experimental verification, thereby complementing
other works probing
the transition between the quantum and the
classical domains\cite{meso}. We conclude by stressing that it should
be worthwhile to compute the results in our example using 
other approaches\cite{reviews} to calculate the quantum arrival 
time distribution, and compare such results with those of the present
paper. Such studies can further motivate the formulation of actual
experiments to decide which particular approach is empirically tenable
for description of the arrival time distribution of quanta in the 
gravitational potential.

\vskip 0.2in
{\bf Acknowledgments}

MMA and AKP acknowledge support of the Senior Research Fellowship from
CSIR, India.  DH is grateful to Paul Davies for stimulating discussions
on this topic and acknowledges the support provided by the Jawaharlal Nehru 
Fellowship.

\end{document}